\documentclass{article}
\usepackage[a4paper, margin=1in]{geometry} %
\usepackage{footnote}
\usepackage{rotating}
\usepackage{graphicx} 
\usepackage{color}
\usepackage{booktabs}
\usepackage{comment}
\usepackage{amsmath}
\usepackage{tabularx}
\usepackage{comment}
\usepackage{hyperref}
\usepackage{caption}
\usepackage{subfigure}
\usepackage{subcaption}
\usepackage{tablefootnote}
\makesavenoteenv{tabular}
\makesavenoteenv{table}
\usepackage{makecell}
\usepackage{authblk}
\usepackage{comment}
\usepackage{natbib}
\hypersetup{
	unicode=false,          
	pdftoolbar=true,        
	pdfmenubar=true,        
	pdffitwindow=false,     
	pdfstartview={FitH},    
	pdftitle={My title},    
	pdfauthor={Author},     
	pdfsubject={Subject},   
	pdfcreator={Creator},   
	pdfproducer={Producer}, 
	pdfkeywords={keywords}, 
	pdfnewwindow=true,      
	colorlinks=true,       
	linkcolor=blue,          
	citecolor=blue,        
	filecolor=magenta,      
	urlcolor=blue           
}

\title{ Evolution of  Coordination Through Institutional Incentives:\\  An Evolutionary Game Theory Approach}

\author[1]{Ndidi Bianca Ogbo \thanks{Corresponding author: \texttt{b.ogbo@tees.ac.uk}}}
\author[1]{Zhao Song}
\author[1]{The Anh Han}
\affil[1]{School of Computing, Engineering and Digital Technologies, Teesside University, United Kingdom}

\begin{document}

\maketitle

\begin{abstract}
There is a broad recognition that commitment-based mechanisms can promote coordination and cooperative behaviours in both biological populations and  self-organised multi-agent systems by making individuals' intentions explicit prior to engagement. Yet their effectiveness depends on sustained compliance supported by institutions, especially in one-off interactions. Despite advances in quantitative studies of cooperation and commitment, most applied analyses and policy debates remain largely qualitative, with limited attention to the allocation of scarce institutional resources between enhancing participation and ensuring commitment compliance. Herein, we develop an evolutionary game-theoretic model that explicitly examines the strategic distribution of a limited budget for institutional incentives, namely rewards or punishments, aimed at these two critical objectives within pre-commitment frameworks. Our findings reveal that a reward-based incentive approach consistently yields greater coordination success than a punishment-based approach, with optimal outcomes arising when resources are appropriately distributed between participation promotion and compliance assurance. These findings offer novel insights for designing institutional incentives to promote broad, coordinated adoption of new technologies.
\end{abstract}

\section{Introduction}

The conflict between self-interest and communal welfare, known as a social dilemma, has become a focal point in recent studies of social interaction and dynamic multi-agent systems (MAS) \cite{key:Sigmund_selfishnes,bloembergen2015evolutionary,fatima2024learning,hanAICOM2022,leung2024promote,PAIVA2018}. Individuals regularly encounter coordination challenges in areas such as climate action, public health, and technology diffusion. Effective coordination in these cases relies  on both rational strategy and the expectation of others' choices. Yet, fostering mutual cooperation among self-interested actors continues to be a pressing challenge across social, economic, and technological systems \citep{key:Sigmund_selfishnes,ostrom1990governing,barrett2016coordination}. A pre-commitment approach, which allows individuals to clarify each other's intentions in advance, offers a solution for addressing coordination challenges  \citep{jennings1993commitments, NBOgbo,HanBook2013,barrett2016coordination,nesse2001evolution}.
Commitment mechanisms restrict an agent’s strategic choices, thereby influencing others’ anticipations and decisions \cite{nesse2001evolution,Singh91intentions_commitments,han2015synergy}. Studies suggest that these mechanisms are evolutionarily robust and effective in promoting cooperation, particularly in one-shot  interactions \citep{NBOgbo,han2022Interface,arvanitis2019,barrett2016coordination}. However, pre-commitment alone is often insufficient. Its success hinges critically on individuals' willingness to participate in the joint commitment, as well as their compliance by participating individuals. Failure to fulfil commitments can destabilise coordination mechanisms, thereby undermining the collective outcomes that such agreements are intended to secure. To ensure the reliability of pre-commitments, robust enforcement mechanisms are necessary  \cite{ostrom2005understanding}. The strategic use of rewards and sanctions serve as key instruments in fostering and sustaining cooperation within structured interactions \citep{Sigmund2001PNAS,han2014cost,han2022Interface,chen2015first,ballietreward2011,dasgupta2025investigating}.

Beyond pre-commitment enforcement, institutional incentives represent a powerful mechanism for promoting prosocial behaviours such as cooperation, coordination and trust  \cite{dong2019competitives,wang2022incentive,wang2023optimization,liu2025evolution,cimpeanu2021}. These systems rely on both positive reinforcement and punitive measures to encourage adherence to collectively beneficial norms. Institutions govern behaviour within various social spheres, such as communities, corporations, and political systems, by regulating behaviour through rule-setting and the use of rewards or penalties to enforce compliance~\cite{ostrom1990governing,ballietreward2011}. Individuals are more inclined to engage in pro-social behaviours that advance collective welfare when enforcement mechanisms are both transparent and perceived as legitimate \citep{sat2021,Duong2021}. Recent research has explored the design of institutional incentives that promote both the initiation of commitments and adherence to them. It has been shown  that incentivising commitment behaviour leads to increased cooperation in spatial public goods games \cite{Flores2024}. Other studies have investigated institutional interventions such as voluntary safety pledges and reputation-based mechanisms, frequently concluding that targeted rewards yield more robust outcomes than punitive measures alone  \citep{han2022Interface,Krellner2025}.

Building on a previous model of commitment-based mechanisms within an evolving population of agents engaged in a cooperation dilemma game \cite{han2022Interface}, we demonstrate that promoting participation is just as crucial as ensuring compliance for sustaining commitment-based coordination. While this previous  model offers valuable insights into evolutionary outcomes, it relies on idealised assumptions and overlooks key institutional factors such as budget constraints \cite{lynch2018fostering,han2018}. To address this limitation, we explicitly model the strategic allocation of limited institutional resources between reward and punishment mechanisms to influence coordination dynamics. Additionally, we focus on an (anti-)coordination problem, which is arguably more challenging to resolve through commitment compared to a cooperation dilemma due to a larger number of potential joint commitments   \cite{NBOgbo,ogbo2022shake}.

In this paper, we investigate how institutional incentives, including both reward and punishment, can be strategically deployed to maximise participation and ensure robust compliance in pre-commitment settings involving coordination problems, with a particular emphasis on socio-technical contexts such as technology adoption. Existing studies demonstrate the efficacy of institutional incentives in promoting cooperation and deterring defection within standard social dilemma frameworks \citep{Duong2021,chen2015first,sigmundetal2001,wang2023emergence,Han2016AAAI}. Yet, these studies primarily focus on post-commitment enforcement and assume binary participation, omitting critical real-world factors such as budget scarcity and allocation constraints faced by institutions.
To address this gap, we explicitly analyse efficient allocation of a finite per capita incentive budget for encouraging commitment participation from players before the game, and for rewarding commitment compliance  and punishment of commitment breaking during the game. A key question is how to distribute this budget to effectively promote coordination success. 

This modelling framework enables a systematic exploration of how varying incentive allocations influence two interconnected behavioural outcomes: individuals’ voluntary participation in pre-commitment and the achievement of mutual coordination. Our results offer policy-relevant insights into how institutions can effectively promote collective action under resource constraints.

\section{Models and Methods}
We begin by examining models that conceptualise technology adoption as a coordination game characterised by asymmetric rewards \cite{NBOgbo,Zhu20033,Chevalier2011}. Our analysis centres on a well-mixed population where individuals can freely choose to pre-commit before engaging in an interaction. Those who commit are then required to cooperate. To investigate strategic technology adoption, we formulate an evolutionary game model involving two investment firms competing in the same market, each deciding which technology to adopt. We first analyse a baseline model that excludes pre-commitment, which is then extended to assess the potential of using commitment-based strategies   for improving coordination between firms.

\begin{table}[t!]
\centering
 \caption{List of parameters in the models.}
\label{tbl:BModel}
    \begin{tabular}{|p{6cm}|c|}
    \hline
    \textbf{Parameters description}  & \textbf{Notation} \\
     \hline \hline
        Cost of investing in high technology, H & $c_H$ \\
      \hline
    Cost of investing in low technology, L & $c_L$\\
    \hline
      Benefit  of investing in high technology, H  & $b_H$\\
   \hline
    Benefit  of investing in low technology, L & $b_L$\\
     \hline
    (Reversed) competitive level of the market & $\alpha$\\
    \hline
     Cost of arranging a commitment & $\epsilon$\\
     \hline
        Per capita budget for incentive & $u$\\
     \hline
       Fraction of per capita budget for rewarding participation  & $\gamma$\\
     \hline
       \end{tabular}
\end{table}

\subsection{Technology Adoption (TD) game} 
We first model technology adoption as a two-player game (strategic interaction) and then introduce the possibility of pre-commitments and institutional incentives to investigate their effect on strategic behaviour.
\label{subsec:PD}
\subsubsection{Two-Player Strategic Interaction in Absence of Pre-commitments}
We consider a non-commitment game in which two firms simultaneously select between two competing technologies in a shared market context \cite{NBOgbo}. Each firm must decide whether to adopt a high-profit, innovative technology (\textbf{H}) or a less profitable, conventional alternative (\textbf{L}). 
The absence of pre-commitment mechanisms implies that coordination is purely strategic, and the payoffs for the row player are presented in the matrix below:
\begin{equation}
\label{eq:PMatrix}
 \bordermatrix{~ & H & L\cr
                  H & \alpha b_H - c_H & b_H - c_H \cr
                  L & b_L - c_L & \alpha b_L - c_L  \cr
                 } =  \ \ \bordermatrix{~ & H & L\cr
                  H &  a & b \cr
                  L & c &  d  \cr
                 }. 
\end{equation}
The parameters $c_L$ and $c_H$ represent investment costs, while $b_L$ and  $b_H$ ($b_L \leq b_H$) denote the corresponding benefits for adopting low or high technology, respectively. The parameter $\alpha \in (0,1)$ captures the competitiveness of the market,  determining the fraction of the market benefit a firm receives when both choose the same technology. A lower $\alpha$ indicates higher competition, making it more advantageous for firms to coordinate on adopting different technologies. For ease of reference, the payoff matrix entries are denoted by $a, \ b, \ c,$ and $d$, as defined above.

Note that, although our model is presented in the context of technology adoption decisions, it is broadly applicable to various other MAS coordination problems, such as strategic investment choices in competitive markets for different products \cite{chevalier2011strategic,Zhu20033}.

\subsubsection{Pre-Commitment Two-Player Strategic Interaction} 
In this framework, commitment is centrally coordinated by a third party prior to gameplay \cite{pereira2017centralized,han2022Interface}. Players may choose to join this arrangement by paying a fixed cost $\epsilon$. Upon joining, they commit to adopting different technologies, with the firm choosing the high-return option (\textbf{H}) providing compensation to the firm choosing the lower-return option (\textbf{L}) \cite{NBOgbo}.
Although players may initially accept the commitment, they might later fail to comply with it. To mitigate this risk, we consider several third-party enforcement strategies, including the provision of rewards for compliance and sanctions for failing to honour an adopted commitment.

In general, a strategy can be defined by a three component, denoted by XYZ,  corresponding to  three  decision points below: 
\begin{enumerate}
    \item Before the game, accept (A) or not (N) to join a commitment to coordinate with each other. That is, $X \in \{A,N\}$. 
    \item if the commitment is not formed (when at least one player does not commit), play the default choice, H or L, in the game. That is, $Y \in \{H,L\}$.  
    \item if the commitment is formed, coordinate (C)  or does not coordinate (D). That is, $Z \in \{C,D\}$.   Playing D means to play the default technology regardless of what other player chooses. This D player will not transfer compensation if the the default choice is H, but will accept compensation if the default choice is L.   
    
\end{enumerate}
Thus, in total, there are 8 possible strategies, summarised in Table 2. In this work, we will perform a comprehensive analysis where the full set of strategies will be included in an evolutionary game analysis (see Methods in Section \ref{section:Methods}).

When two players accept a commitment is formed. 
If both play C, they get the same payoff $(b+c-\epsilon)/2$. 
If one plays C and the other plays D, the latter would choose their default technology in the TD game, and will not transfer compensation if the this default choice is H.
For example, when AHC plays AHD, latter chooses H in the game while the former coordinates and play L. AHC  receives $c-\epsilon/2$ while AHD $b+\epsilon/2$, as there is not compensation.  
When AHC plays ALD, latter chooses L in the game while the former coordinates and play H and transfer the compensation. They both get $(b+c-\epsilon)/2$.

When AHD plays AHD, both play H and receive $a$, still sharing the cost $\epsilon$, thereby receiving: $a-\epsilon/2$. 
When AHD plays ALC, the former choose H and the letter plays C. Yet the former does not compensate. Thus, AHD receives $b-\epsilon/2$ and ALC receives $c-\epsilon/2$.
When AHD plays ALD, they choose H and L respectively but the former does not compensate. AHD receives $b-\epsilon/2$ and ALD $c-\epsilon/2$.

When at least one player does not accept, no agreement is formed. The players use the choice in  absence of a commitment, see Table 1. 
For example, when AHC plays NHC, both plays H and receive $a$.

\begin{table}
\begin{center}
\begin{tabular}{ p{1.2cm}||p{1.7cm}|p{1.8cm}|p{2cm} }
 \hline
Strategies & Accept commitment to coordinate? & Play H in absence of commitment? & Coordinate in presence of commitment? \\
 \hline
 AHC   & Yes    &  Yes  &  Yes \\
 AHD &   Yes  & Yes   &No\\
 ALC &Yes & No&  Yes\\
 ALD    &Yes & No&  No\\
 NHC &   No  & Yes&Yes\\
 NHD& No  & Yes   &No\\
 NLC& No  & No&Yes\\
  NLD& No  & No&No\\
 \hline
\end{tabular}
\caption{The eight strategies with commitment formation.}
\end{center}
\label{table:eight-strategies1}
\end{table}

\begin{equation} 
 \label{payoff_matrix_share_cost}
\bordermatrix{~ & \text{AHC}  & \text{AHD}  & \text{ALC} & \text{ALD} & \text{NHC} & \text{NHD} & \text{NLC} & \text{NLD} \cr
  \text{AHC}&  \frac{b+c-\epsilon}{2} & c  -\frac{\epsilon}{2} & \frac{b+c-\epsilon}{2} &\frac{b+c-\epsilon}{2} & a & a & b & b  \cr
\text{AHD} &b-\frac{\epsilon}{2}& a -\frac{\epsilon}{2}& b-\frac{\epsilon}{2}  & b-\frac{\epsilon}{2}  & a & a & b & b  \cr
 \text{ALC} &\frac{b+c-\epsilon}{2}& c-\frac{\epsilon}{2} & \frac{b+c-\epsilon}{2}& \frac{b+c-\epsilon}{2}  & c  & c & d & d   \cr
 \text{ALD} &\frac{b+c-\epsilon}{2} & c-\frac{\epsilon}{2} & \frac{b+c-\epsilon}{2}  & d -\frac{\epsilon}{2} & c  & c & d & d   \cr
 \text{NHC} &a & a &b  & b  & a & a & b & b  \cr
 \text{NHD} &a & a &b  & b  & a & a & b & b  \cr
 \text{NLC} &c & c  &d  & d  & c & c & d & d  \cr
 \text{NLD} &c  & c  &d  & d  & c & c & d & d  \cr
 }
\end{equation}

\subsubsection{Institutional reward and punishment in absence of noise}
We assume that there is a per capita budget $u$ available for providing incentives. A fraction  of the budget, $\gamma u$ ($0 \leq \gamma \leq 1$), is used for rewarding those who are  willing to participate in a commitment (i.e.  AXY players for $X \in \{H,L\}$ and $Y \in \{C,D\}$), increasing the chance a commitment being formed. The remaining budget, i.e. $(1-\gamma) u$, is used for rewarding commitment compliant players (i.e. AHC and ALC players; and ALD players in interactions with AHC and ALC, as they still appear to comply) or punishing non-compliant ones (i.e. AHD players; and ALD players when interacting with AHD and ALD, as they still appear to comply when interacting with AHC and ALC players). When $\gamma = 0$, it means the budget is used only for incentivising commitment compliant  behaviours (i.e., \textit{pure reward} and \textit{pure punishment} scenarios). 
As we consider reward and punishment separately, without loss of generality, we assume that all the incentives described above are equally cost efficient, where the incentive recipient's increased or decreased amount (corresponding to reward and punishment, respectively) equals the institution's cost.  

When a per capita budget $u$ is available to \textit{reward commitment compliant behaviours}, with a fraction $\gamma$ of it being used for rewarding participation, the payoff matrix reads

\begin{equation} 
 \label{payoff_matrix_share_cost1}
\bordermatrix{~ & \text{AHC}  & \text{AHD}  & \text{ALC} & \text{ALD} & \text{NHC} & \text{NHD} & \text{NLC} & \text{NLD} \cr
  \text{AHC}&  \frac{b+c-\epsilon}{2} + u  & c  -\frac{\epsilon}{2} + u&  \frac{b+c-\epsilon}{2} + u& \frac{b+c-\epsilon}{2} + u&  a + \gamma u& a + \gamma u& b+ \gamma u & b + \gamma u \cr
\text{AHD} &b-\frac{\epsilon}{2}+ \gamma u& a -\frac{\epsilon}{2}+ \gamma u& b-\frac{\epsilon}{2}+ \gamma u& b-\frac{\epsilon}{2}+ \gamma u &  a + \gamma u& a + \gamma u & b+ \gamma u & b+ \gamma u  \cr
 \text{ALC} & \frac{b+c-\epsilon}{2} + u  & c  -\frac{\epsilon}{2} + u & \frac{b+c-\epsilon}{2} + u &\frac{b+c-\epsilon}{2} + u   &  c + \gamma u&  c + \gamma u& d + \gamma u& d + \gamma u  \cr
 \text{ALD} & \frac{b+c-\epsilon}{2} + u & c-\frac{\epsilon}{2}+ \gamma u &  \frac{b+c-\epsilon}{2} + u&  d -\frac{\epsilon}{2} + \gamma u&  c + \gamma u&  c + \gamma u& d + \gamma u& d+ \gamma u   \cr
 \text{NHC} &a & a &b  & b  & a & a & b & b  \cr
 \text{NHD} &a & a &b  & b  & a & a & b & b  \cr
 \text{NLC} &c & c  &d  & d  & c & c & d & d  \cr
 \text{NLD} &c  & c  &d  & d  & c & c & d & d  \cr
}.
\end{equation}

Similar to the reward case, the payoff matrix  in the punishment setting reads:

\begin{equation}
\label{payoff_matrix_share_cost2}
\bordermatrix{~ & \text{AHC} & \text{AHD} & \text{ALC} & \text{ALD} & \text{NHC} & \text{NHD} & \text{NLC} & \text{NLD} \cr
\text{AHC} & \lambda_1 & \lambda_2 & \lambda_1 & \lambda_1 & a + \gamma u & a + \gamma u & \lambda_3 & \lambda_3 \cr
\text{AHD} & \lambda_4 & \lambda_5 & \lambda_4 & \lambda_4 & a + \gamma u & a + \gamma u & \lambda_3 & \lambda_3 \cr
\text{ALC} & \lambda_1 & \lambda_2 & \lambda_1 & \lambda_1 & c + \gamma u & c + \gamma u & d + \gamma u & d + \gamma u \cr
\text{ALD} & \lambda_1 & \lambda_6 & \lambda_1 & \lambda_7 & c + \gamma u & c + \gamma u & d + \gamma u & d + \gamma u \cr
\text{NHC} & a & a & b & b & a & a & b & b \cr
\text{NHD} & a & a & b & b & a & a & b & b \cr
\text{NLC} & c & c & d & d & c & c & d & d \cr
\text{NLD} & c & c & d & d & c & c & d & d \cr
},
\end{equation}
where,  for the sake of notational  brevity, we use the following abbreviations  $\lambda_1 = \frac{b + c - \epsilon}{2} + \gamma u$, $\lambda_2 = c - \frac{\epsilon}{2} + \gamma u$, $\lambda_3 = b + \gamma u$, $\lambda_4 = b - \frac{\epsilon}{2} + (2\gamma - 1) u$, $\lambda_5 = a - \frac{\epsilon}{2} + (2\gamma - 1) u$, $\lambda_6 = c - \frac{\epsilon}{2} + (2\gamma - 1) u$, and $\lambda_7 = d - \frac{\epsilon}{2} + (2\gamma - 1) u$.

\subsection{Evolutionary Dynamics in Finite Population}
\label{section:Methods}
This study employs evolutionary game theory methods for finite populations \cite{Nowak20044,Imhof20055}, combining numerical simulations with theoretical analysis. Let $Z$ denote the population size. Within finite populations, individual payoffs are understood as measures of \emph{fitness} or social \emph{success}. Evolutionary change occurs through social learning, in which agents tend to imitate those with higher success \cite{key:Sigmund_selfishnes,Hofbauer19988}.
We implement social learning using the pairwise comparison rule \cite{key:Sigmund_selfishnes,Traulsen20066}, a well-established framework in evolutionary dynamics. Under this rule, an individual $A$ with fitness $f_A$ adopts the strategy of another individual  $B$ with fitness $f_B$ with probability $p$ given by the Fermi function, $p_{A, B}=\left(1 + e^{-\beta(f_B-f_A)}\right)^{-1}.$

Here $\beta$ represents the strength of imitation or selection intensity, indicating how sensitive individuals are to fitness differences in the imitation process. At $\beta=0$, strategy adoption is random (neutral drift), while higher values of $\beta$ lead to more deterministic imitation favouring fitter individuals.

Without mutation or exploration, evolution leads to monomorphic states that are absorbing under imitation, meaning the population cannot escape once such a state is reached. To allow for continued evolution, we introduce a mutation mechanism: with a small probability, individuals may randomly adopt a different strategy, independent of imitation.
When mutation rates are sufficiently low, the population typically contains no more than two strategies at a time. This allows the evolutionary process to be modelled as a Markov chain, with each state representing a monomorphic population and transition probabilities determined by mutant fixation likelihoods \cite{Hofbauer19988,Imhof20055}. The resulting Markov chain possesses a stationary distribution that characterizes the long-run proportion of time the population spends in each monomorphic state. Notably, this approximation remains valid even beyond the strict limit of infinitesimally small mutation or exploration rates \cite{Traulsen20066,Hauert20077,key:Hanetall_AAMAS2012,key:sigmund2010}.

Let $\pi_{ij}$ represent the payoff obtained by strategist $i$ in each pairwise interaction with strategist $j$, as defined in the payoff matrices. Suppose there are at most two strategies in the population, say, $  {x}$ individuals using $i$ ($0\leq   {x} \leq Z$) and ($Z-  {x}$)  individuals using $j$. Thus the average payoff of the individual that uses $i$ or $j$ can be written respectively as follows;
\begin{equation} 
\label{eq:PayoffA}
\begin{split} 
\Pi_i(  {x}) &=\frac{(  {x}-1)\pi_{ii} + (Z-  {x})\pi_{i,j}}{Z-1},
\\
\Pi_j(  {x}) &=\frac{  {x}\pi_{j,i} + (Z-  {x}-1)\pi_{j,j}}{Z-1}.
\end{split}
\end{equation} 

Now, the probability to change the number $k$ of agents using strategy A by $\pm$ one in each time step can be written as \citep{traulsen2006} 
\begin{equation} 
T^{\pm}(k) = \frac{N-k}{N} \frac{k}{N} \left[1 + e^{\mp\beta[\Pi_A(k) - \Pi_B(k)]}\right]^{-1}.
\end{equation}
The fixation probability of a single mutant with a strategy A in a population of $(N-1)$ agents using B is given by \citep{traulsen2006,key:novaknature2004}
\begin{equation} 
\label{eq:fixprob} 
\rho_{B,A} = \left(1 + \sum_{i = 1}^{N-1} \prod_{j = 1}^i \frac{T^-(j)}{T^+(j)}\right)^{-1}.
\end{equation} 

Considering a set  $\{1,...,q\}$ of different strategies, these fixation probabilities determine a transition matrix $M = \{T_{ij}\}_{i,j = 1}^q$, with $T_{ij, j \neq i} = \rho_{ji}/(q-1)$ and  $T_{ii} = 1 - \sum^{q}_{j = 1, j \neq i} T_{ij}$, of a Markov Chain. The normalized eigenvector associated with the eigenvalue 1 of the transposed of $M$ provides the stationary distribution described above \citep{key:imhof2005}, describing the relative time the population spends adopting each of the strategies.

\section{Results}
We present results from numerical simulations (using Methods described in Section 2) that explore the extent to which targeted institutional incentives, including reward for participation, reward for commitment compliance and punishment for non-compliance, influence pre-commitment and mutual coordination among agents.

\subsection{Effect of Institutional Incentive Budget on Coordination  and Strategic Behaviours}
Figures 1 and 2 investigate how changes in per capita institutional incentive budget ($u$) affect both the total coordination level (denoted as \textbf{To-C}) and how strategies perform within the population in presence of all other strategies (see Table 2). 

\begin{figure*}[htbp]
\centering
\includegraphics[width=0.9\textwidth]{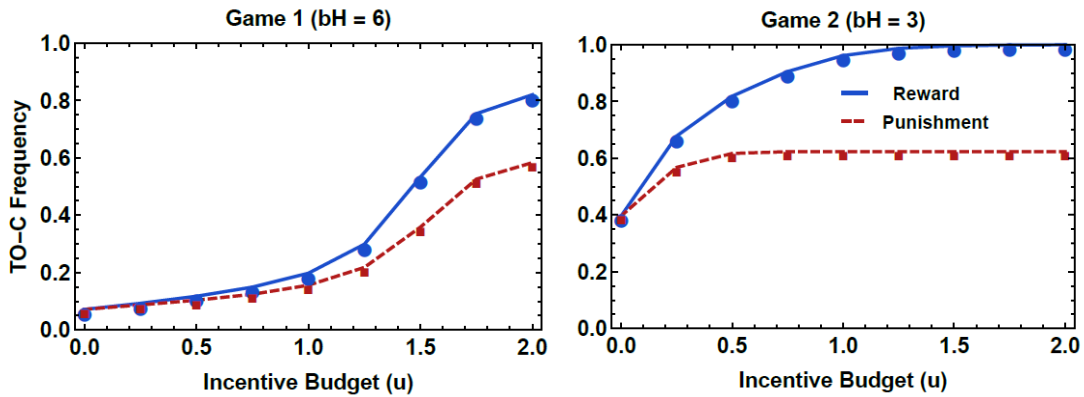}  
\caption{Frequency of Total coordination (To-C) as a function of $u$ (incentive budget) under institutional reward and punishment. Parameters: in all panels: $c_H = 1$, $c_L = 1$, $b_L = 2$ (i.e. $c = 1$); Game 1 (left): $b_H = 6$ (i.e. $b = 5$), Game 2 (right): $b_H = 3$ (i.e. $b = 2$). Other parameters: $\beta = 0.1$; $\alpha =0.5$; $\epsilon = 1$; $\gamma =0$; population size $N = 100$.} 
\label{fig:fig1}
\end{figure*}

 \begin{figure}[!ht]
\centering
\includegraphics[width=0.9\textwidth]{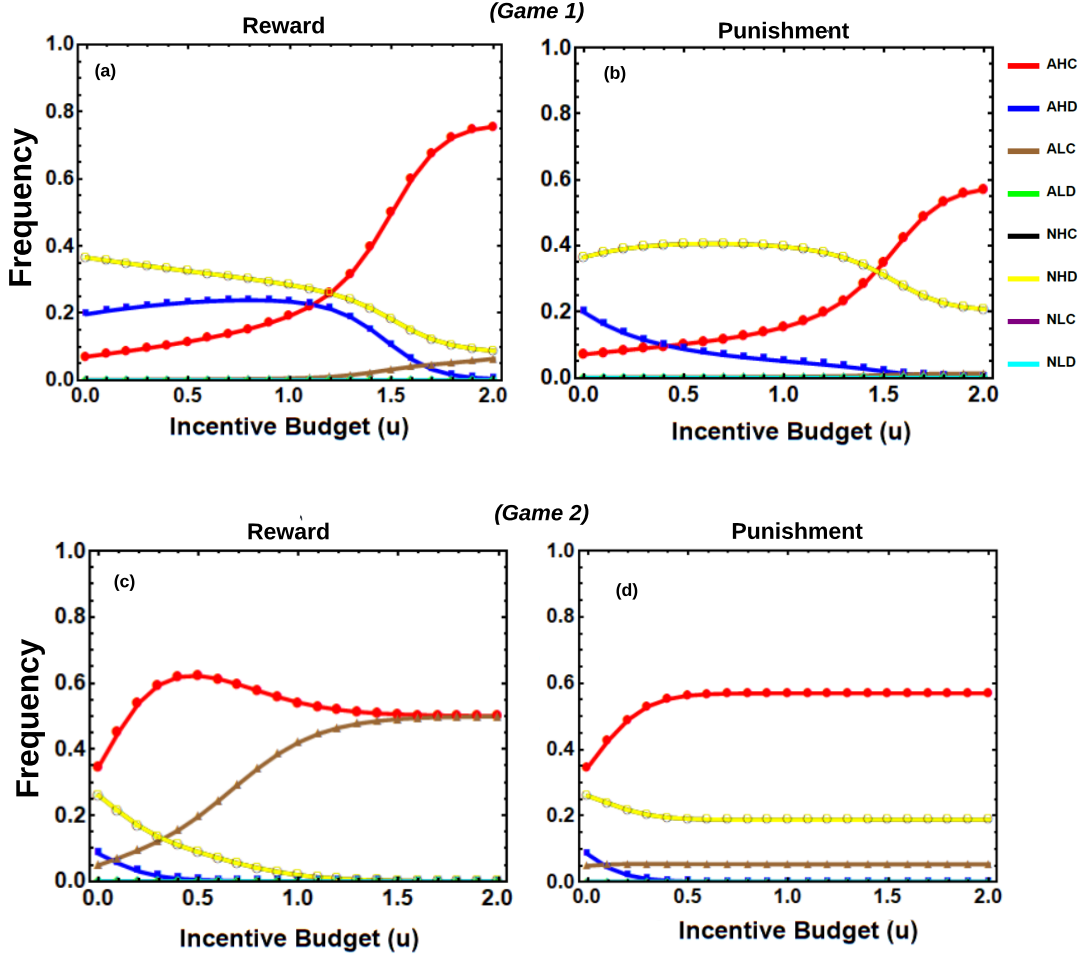} 
\caption{\textbf{Frequency of 8 strategies (see Table 2) as a function of $u$ (incentive budget) under institutional reward and punishment approaches.} Parameters: in all panels: $c_H = 1$, $c_L = 1$, $b_L = 2$ (i.e. $c = 1$); Game 1 (top row): $b_H = 6$ (i.e. $b = 5$); Game 2 (bottom row): $b_H = 3$ (i.e. $b = 2$). Other parameters: $\beta = 0.1$; $\alpha =0.5$; $\epsilon = 1$; $\gamma =0$; population size $N = 100$.} 
\label{fig:fig18}
\end{figure}

In all incentive scenarios, Figure 1 demonstrates that To-C rises monotonically with $u$, suggesting that higher institutional budgets consistently improve cooperation and coordination. Nevertheless, over the whole budget range, reward-based incentives consistently yield greater To-C values than punishment-based incentives. The benefits of rewards are especially noticeable in low-budget regimes and in Game 2, where individual incentives to coordinate are less due to the less favourable coordination environment. Positive reinforcement has a high marginal effectiveness in promoting commitment-based coordination, as seen by the significant To-C benefits that arise from even modest allocations to incentives

These overall variations are explained by the disaggregated data in Figure 2, where we show the frequencies of the eight strategies (see Table 2) for varying the incentive budget ($u$). In reward-based mechanisms, raising the budget  leads to a quick increase in commitment compliant strategies (AHC and ALC) while decreasing non-compliant strategies (AHD and ALD). This trend implies that rewards effectively convert agents who join commitments but might otherwise deviate into conditional  coordinators. Strategies such as NHC and NLC, are also declining as incentives for commitment participation become more attractive. Fully non-committing strategies like the NHD and NLD, are the most resistant to reward-induced change, but their prevalence drops significantly at higher budgets.

These findings are consistent with theoretical and empirical research indicating that sanctions can deter participation when compliance is uncertain \cite{key:Hauert2007,Rand2011,FehrAmerRev2000}. Negative incentives may also overshadow intrinsic pro-social drive and reduce views of institutional legitimacy \cite{FehrFalk2002,ballietreward2011,ostrom2005understanding}. Overall, rewards improve the efficiency with which institutional funds are translated into coordination, resulting in increased To-C at lower thresholds and shifting strategic distribution towards AHC and ALC. Punishments slow down coordination, increase strategic heterogeneity, and perform less well across situations.

\begin{figure}[!ht]
\centering
\includegraphics[width=0.9\textwidth]{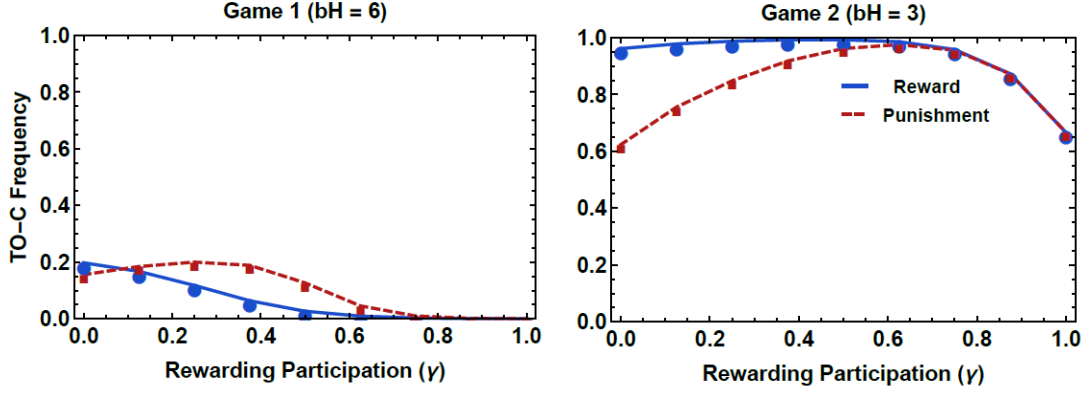}  
\caption{\textbf{Frequency of Total Coordination (To-C) as a function of the fraction of per capital budget used for encouraging  participants ($\gamma$) under institutional reward and punishment.} Parameters: in all panels: $c_H = 1$, $c_L = 1$, $b_L = 2$ (i.e. $c = 1$); Game 1 (left): $b_H = 6$ (i.e. $b = 5$); Game 2 (right): $b_H = 3$ (i.e. $b = 2$). Other parameters: $\beta = 0.1$; $\alpha =0.5$; $\epsilon = 1$; $u = 1$; population size $N = 100$.} 
\label{fig:fig2}
\end{figure}

\begin{figure}[!ht]
\centering
\includegraphics[width=\textwidth]{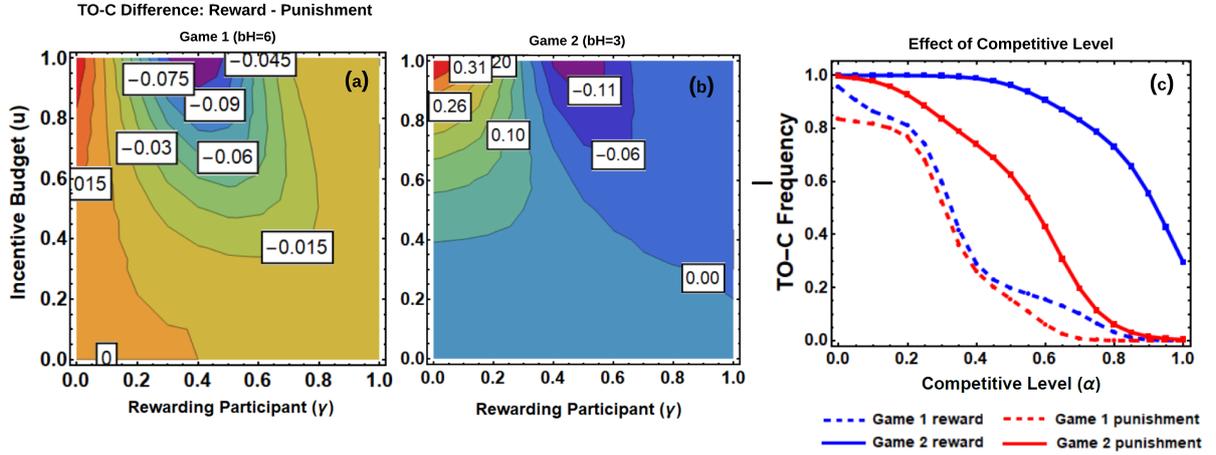}  
\caption{\textbf{Panels (a) and (b): we show the difference of the total coordination levels achieved through reward and punishment, as a function of $\gamma$ and $u$ ($\alpha =0.5$).  Panel (c): We show the total coordination levels achieved via reward vs punishment for  varying the competitive level of the market $\alpha$ ($u=1$, $\gamma=0$)}. Parameters: in all panels: $c_H = 1$, $\beta = 0.1$, $\epsilon = 1$,  $N = 100$, $c_L = 1$, $b_L = 2$ (i.e. $c = 1$); Game 1: $b_H = 6$ (i.e. $b = 5$), Game 2: $b_H = 3$ (i.e. $b = 2$). }
\label{fig:fig3}
\end{figure}

\subsection{Budget Fraction for Participation vs Coordination Outcomes}
In Figure 3, we analyse how adjustments in the budget share allocated to rewarding participation ($\gamma$) influence the frequency of total coordination (To-C) in both reward and punishment cases.
To-C stays low across all values of rewarding participation ($\gamma$) in Game 1, where baseline incentives for coordination are inherently strong. With increasing $\gamma$ under rewards, To-C monotonically declines until it approaches zero when the majority of resources are allocated to participation.
Across all $\gamma$ levels, punishment incentives are minimal. They only slightly increase around $\gamma \approx 0.2$ before gradually decreasing. The consistently low To-C levels suggest that reallocating resources towards participation at the expense of compliance enforcement gives little advantage and may even impair performance in settings that are already supportive of coordination.
With less strong intrinsic incentives, patterns are very different in Game 2. Rewards remain close to the maximum To-C $\gamma \approx 1.0$ throughout a wide range, only decreasing when $\gamma > 0.9$. With increasing sensitivity to allocation balance, punishments get better with $\gamma$, peaking at $\gamma \approx 0.6$ before suddenly declining.

According to the results, the underlying coordinating environment determines how well institutional resources are allocated to participation rewards. Higher $\gamma$ values did not improve coordination in Game 1, where baseline incentives for coordination are naturally strong. This suggests that placing too much emphasis on participation could take resources away from compliance enforcement and lower overall performance \cite{Duong2021}.
Game 2 incentives, on the other hand, maintained excellent coordination over a wide range of allocations, but punishments reached lower maxima and needed to be precisely calibrated to approach similar levels. These findings imply that while reward-based mechanisms are typically more resilient to changes in allocation \cite{sat2021,Hauert20077}, their ideal setup needs to be customised for contextual incentive structures, and that even in favourable environments, high participation rewards are not enough to sustain coordination.

\begin{figure}[ht!]
\centering
\includegraphics[width=\textwidth]{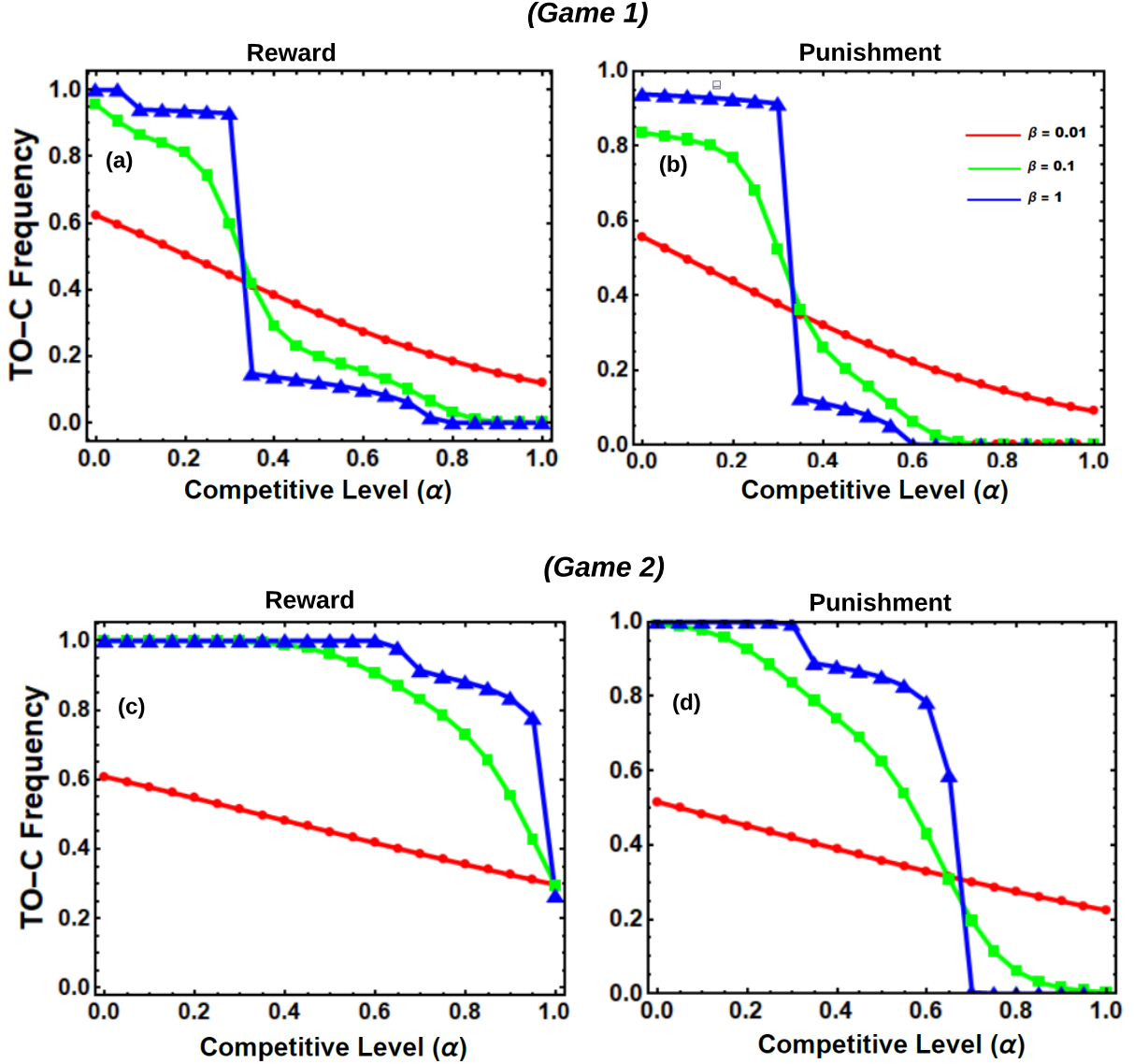}  
\caption{\textbf{We show the total coordination levels achieved via institutional reward (left column) vs punishment (right column) for  varying the competitive level of the market for different intensities of selection ($\beta$)}. Parameters: in all panels: $\epsilon = 1$,  population size $N = 100$, $c_L = 1$, $b_L = 2$ (i.e. $c = 1$), $u=1$, $\gamma=0$. Game 1 (top row): $b_H = 6$ (i.e. $b = 5$); Game 2 (bottom row): $b_H = 3$ (i.e. $b = 2$).}
\label{fig:fig7}
\end{figure}

\subsection{Joint Effects of Incentive Allocation and Market Competition on Coordination}

Figure 4 (a)-(b) compare the total coordination of reward and punishment incentives across participation reward ($\gamma$) and incentive budgets ($u$). Higher To-C under rewards is indicated by positive values, while higher To-C under punishments is indicated by negative values. 
In Game 1 (panel a), punishments typically perform better than rewards especially at moderate allocations. The biggest advantage of about -0.075 occurs between $u \approx 0.9\, \text{and} \, \gamma \approx 0.3$, while rewards only produce slight improvements in specific areas such  $u \approx 0.6\, \text{and} \, \gamma \approx 0.1$ with a value of +0.015.
Game 2 (panel b), in contrast shows a wider dominance of rewards, with the most noticable advantage of 0.31 at $u \approx 1.0\, \text{and} \, \gamma \approx 0.1$, and consistent favourable effects at intermediate allocations (for instance, +0.10 at $u \approx 0.7\, \text{and} \, \gamma \approx 0.3$).
The severity of punishments slightly increases, reaching a high value at -0.11 when $u \approx 0.9\, \text{and} \, \gamma \approx 0.6$. 
These findings align with figure \ref{fig:fig2} result, where reward incentives maintain more coordination over a longer range and TO-C reaches its highest levels at intermediate $\gamma$ values in Game 2. 
The results collectively show that while sanctions work better in resource-rich environments with balanced compliance enforcement, rewards offer resilience under low-to-moderate allocations.
This synergy demonstrates the potential of hybrid incentive schemes, which combine targeted punishments to discourage persistent defection with rewards to ensure cooperation \cite{Xu2021Synergy,Andreoni2003Carrot}.

Figure 4(c) illustrates how coordination decreases as $\alpha$ rises, with high $\alpha$ values denoting less competitive settings and low $\alpha$ values (0–0.5) indicating intense market competition. Throughout the majority of $\alpha$ values in Game 1, reward incentives consistently outperform punishments; but, in less competitive environments, the gap narrows $\alpha > 0.5$. In Game 2, punishments drastically decrease after $\alpha > 0.4$, while reward incentives maintain a high To-C even in the face of fierce competition $\alpha \leq 0.7$. These findings collectively reveal that budget size, allocation balance, and competitive environment all affect how effective hybrid or reward-oriented methods are, with rewards demonstrating a higher level of resilience to fierce market competition.

\subsection{Effect of Market Competition and Selection Intensity on Coordination }

Figure 5 illustrates how the interplay  between selection intensity ($\beta$) and competitive level ($\alpha$) affects coordination outcomes in the population, in the presence of either institutional reward or punishment. In both games, rewards are preferred over punishments in highly competitive market conditions (i.e. $\alpha \leq 0.5$), and the advantage grows under strong selection ($\beta = 1$). 
Rewards maintain a high To-C under these circumstances until $\alpha$ gets close to 0.4 or 0.6, but punishments lead to a faster decrease. Although both approaches become less efficient as competition declines ($\alpha > 0.5$), rewards continue to be superior.

 An important interpretation of these results is the role of selection intensity ($\beta$) when examining variations across populations (e.g. regions, cultures). In line with \cite{traulsen2010human,Duong2021,GokhalePNAS2010,randUltimatum}, $\beta$ can be  understood as a measure of the extent to which agents’ strategy updates are conditioned by payoff differentials. From a cross-population perspective, $\beta$ may vary systematically with cultural norms, institutional trust, and the efficacy of social learning processes \cite{Duong2021,GokhalePNAS2010}. In high $\beta$ contexts, representing for example fast-moving societies \cite{zimmaro2024emergence}, even minor payoff disparities can prompt swift behavioural changes, increasing the responsiveness of coordination equilibria to slight modifications in incentive frameworks. Behaviour sensitivity to reward differences is reduced in low $\beta$ situations, which are characterised by weak institutional reach or high levels of uncertainty. As a result, more or longer-lasting incentives are needed to produce coordination outcomes comparable to those in high $\beta$ contexts.  This interpretation is supported by Figure 5's patterns, which demonstrate that under reward-based incentives, stronger selection increases behavioural responsiveness while decreasing stability under punitive regimes. The external validity and policy relevance of the model are improved by connecting $\beta$ to observable socio-institutional features.

\section{Discussion and Conclusion}
It has been proposed that the human capacity for commitment evolved through natural selection as an adaptive mechanism to facilitate and stabilise cooperation in strategic or uncertain social contexts \cite{nesse2001evolution,Akdeniz2021}. Establishing mutual commitments prior to engagement increases the likelihood of successful collective  outcomes \cite{NBOgbo,han2022Interface,IronsChapter2001,Cherry2013}, as it allows individuals to clarify intentions and coordinate expectations before embarking on potentially costly courses of action. However, in the absence of third-party enforcement, commitments are typically self-imposed, leaving individuals with the discretion to renege on agreements. The prevalence of commitment violations observed in empirical and experimental research \cite{han2022Interface,sasaki2015commitment,Dannenberg2016,Nguyen2019} has led to increasing interest in identifying which institutional mechanisms, particularly those involving rewards or sanctions, best uphold the cooperation-promoting benefit of commitments.

Using numerical simulations and a mathematical two-stage game-theoretic framework as a foundation, our study explored how institutional budget ($u$) and its distribution between encouraging  participation and enforcing compliance ($\gamma$) affect coordination outcomes (To-C) in an evolving population. Our findings show a number of strong trends that connect institutional design decisions to both individual behavioural composition and collective coordinating outcomes. Increasing $u$ improves To-C in both systems,  although rewards work better than penalties across all budgets. While sanctions leave residual defection, our finding  shows that rewards more successfully turn fake compliant players (i.e. AHD and ALD) into compliant coordinators (i.e. AHC and ALC) and decrease non-participatory strategies' frequencies.

Our results  show that coordination peaks when a sufficient amount of the incentive budget is  used to encourage participation, especially  in scenarios  where performance is negatively impacted by overinvesting in either compliance or participation. In contrast to punishments, which are more allocation-sensitive, rewards maintain high levels of cooperation over a wide  range of $\gamma$. This trade-off points to the possible advantages of hybrid schemes, which lessen the sensitivity of strictly punitive systems by securing participation through rewards and discouraging persistent defection through targeted punishments \cite{Xu2021Synergy,Hilbe2010Evolution,Andreoni2003Carrot}. Majority of the rewarding participants' ($\gamma$) parameter space is dominated by reward-based mechanisms. Their benefit is greatest at low to moderate budgets, where the marginal gains from more institutional resources are highest. The effectiveness of rewards in converting tight budgets into significant coordination gains is seen by this trend. 
We further showed that  high  market competition levels  and a strong intensity of selection  favour the use of rewards, since these circumstances make behaviour more receptive to rewards and make punitive systems less stable. This allows for a cross-population  view by interpreting $\beta$ as a proxy for governance capacity, institutional quality, and cultural responsiveness to incentives \cite{Bowles2012Economic,Henrich2010Markets,Acemoglu2012Why,North1990Institutions}. High $\beta$ environments have the ability to quickly transform reward-based interventions into coordination advantages.

From a practical standpoint, these findings align with recommendations, e.g. from the UK Government’s Technology Adoption Review~\cite{DSIT2025}, which highlights the role of targeted financial incentives in addressing coordination failures and accelerating technology adoption. The report underscores that, despite demonstrable effectiveness of technology, widespread adoption of technology is hindered by misaligned individual incentives and weak institutional backing \cite{rogers2003diffusion}. Moreover, the work in \cite{AlfaroSerrano2021} provides empirical support for this perspective through a comprehensive systematic review of 80 studies across 45 countries. The results of the finding show that financial incentives can help minimise barriers and promote firm-level coordination of technology adoption. Additionally, findings in the context of Ghana shows that the joint application of financial incentives and peer endorsement significantly elevates and sustains technology adoption rates \cite{Riley2025,rogers2003diffusion}. In Pakistan, \cite{Atkin2017} demonstrate that subsidising mechanized cutting technology within a production cluster facilitates adoption by reducing the costs associated with being an early mover, particularly when adoption is coordinated among neighbouring firms. To address this gap, both clear regulatory frameworks and behavioural mechanisms are needed to reward early participation and reduce commitment barriers, thereby highlighting the importance of structured incentives in driving coordinated action~\cite{DSIT2025}.

Future research should explore extended models that include network-structured interactions and agent heterogeneity. Cooperation trajectories can be strongly impacted by heterogeneity in agent preferences, risk attitudes, and enforcement sensitivity \cite{perc2017statistical,SantosPNAS2011,ogbo2022shake}. Similarly, compared to well-mixed populations, structured populations with spatial or community-level clustering frequently display distinct stability qualities, which may have consequences for the best possible incentive targeting \cite{Szabo2007,wang2023optimization}. 
Coordination resilience and cost-effectiveness may be further enhanced by adaptive allocation systems, in which organisations adjust participation and compliance budgets in reaction to observed behaviours. 
Hybrid incentive schemes, which combine the deterrent effect of selective punishments with trust-building and engagement benefits of rewards, need further investigation, as previous research indicates that mixed strategies can outperform purely positive or purely negative approaches in maintaining cooperation \cite{Xu2021Synergy,chen2015first}. Finally, validating these theoretical insights through behavioural experiments or empirical field data would enhance our understanding of their practical relevance and guide policy responses to collective action issues in areas such as climate change mitigation, health, and technology adoption.

\section*{Acknowledgment}
T.A.H. is supported by EPSRC (grant EP/Y00857X/1)
\section*{Author contributions}
N.B.O and T.A.H conceived and designed the study; all authors performed research; all authors analyzed results and wrote the manuscript.

\section*{Competing interest} Authors declare that they have no conflict of interest.

\section*{Data availability}
No datasets were generated or analysed during the current study. 
\clearpage
\bibliographystyle{unsrt}
\bibliography{reference}

\end{document}